\documentclass[prd,showpacs,preprintnumbers,nofootinbib,aps]{revtex4}
\usepackage{subeqnarray}
\usepackage{epsfig,amsmath,amssymb}
\usepackage{mathrsfs}
\usepackage[usenames,dvipsnames]{color}
\usepackage[pagebackref=false, colorlinks=false]{hyperref}
\definecolor{redish}{rgb}{0.7,0.2,0.0}  
\definecolor{bluish}{rgb}{0.2,0.5,0.8}
\hypersetup{linkcolor=redish,          
                  citecolor=blue,        
                  filecolor=magenta,      
                  urlcolor=bluish}          
\DeclareFontFamily{U}{rsfs}{}         
\DeclareFontShape{U}{rsfs}{m}{n}{<5> rsfs5 <6><7> rsfs7          %
  <8><9><10><10.95><12><14.4><17.28><20.74><24.88> rsfs10}{}     %
\DeclareMathAlphabet{\mathfs}{U}{rsfs}{m}{n}                     %

\newcommand{\ba}{\nopagebreak[3]\begin{eqnarray}}
\newcommand{\ea}{\end{eqnarray}}
\newcommand{\bii}{\begin{itemize}}
\newcommand{\eii}{\end{itemize}}

\def \m{\mu}
\def \n{\nu}
\def \r{\rho}

\begin{document}

\title{The Relativistic Point Charge Revisited : Novel Features }
\author{Anarya Ray}
\email{ronanarya9988@gmail.com}
\affiliation{Department of Physics, Presidency University, Kolkata, India}
\author{Parthasarathi Majumdar and Zahid Ansari}
\email{bhpartha@gmail.com}
\email{zahidansari8017@gmail.com}
\affiliation{School of Physical Sciences, Indian Association for the Cultivation of Science, Kolkata 700032, India}
\pacs{04.70.-s, 04.70.Dy}

\begin{abstract}

A relativistically covariant energy-momentum tensor for electromagnetic radiation due to an arbitrarily moving point charge is derived for the first time (as far as we know), directly from the covariant  field strengths extracted from the Lienard-Wiechert potentials.  This energy momentum tensor yields results for the radiant angular power distribution, in full agreement with results derived in a frame-dependent manner in standard texts of classical electrodynamics. This tensor is then used to present a full derivation, not available in standard texts, of the energy-momentum and orbital angular momentum of a relativistic point charge. Radiation backreaction is turned on and the system reanalyzed Lorentz-covariantly, including effects of mass renormalization. This leads us to reiterate earlier conclusions regarding the inherent inadequacy of classical Maxwell electrodynamics. The Abraham-Lorentz equation is derived en passant in appropriate limits without requiring any extraneous structural artifact, and the Landau-Lifschitz proposal for modification of the theory is also critically reviewed.
\end{abstract}
\maketitle

\section{Introduction}

A manifestly Lorentz-invariant formulation of Maxwell electrodynamics (with and without sources), based on {\it physical} potentials (without using any gauge fixing) rather than electric and magnetic field strengths, has been given recently in a companion paper \cite{mr}. The key physical input used in this formulation is the ubiquitous existence of electromagnetic waves, which dictates the nature of the starting equations for the physical potentials. With well-known definitions of the electric and magnetic field strengths in terms of the potentials, first given by Kelvin and Maxwell in the 19th century, the standard textbook Maxwell equations (in vacuum, with and without sources) for ${\vec E}$ and ${\vec B}$ fields are obtained easily. However, the simplicity of the foundational equations for the potentials makes them the prime candidates for solution in various physical circumstances; the field strengths in each case are easily determined by computing a few spacetime partial derivatives of the potentials. The logical self-consistency of our approach has been discussed in detail in ref. \cite{mr}.    

The obvious application of our formulation is to the theory of electromagnetic radiation. In this paper, we develop a manifeslty relativistically covariant formalism \cite{einst} to derive an energy-momentum tensor of the radiation potentials due to an arbitrarily moving relativistic point charge. This Lorentz-covariant energy-momentum tensor has not appeared in the excellent textbooks by Landau and Lifschitz \cite{ll-ctf} and by Jackson \cite{jack}. We are not aware of other textbooks in which it has appeared, if at all.

While the energy and momentum of radiation due to a {\it non-}relativistic point charge are easily obtained from the full energy-momentum tensor for a relativistic charge in the appropriate approximation, with the angular integrations being reduced to quadratures, the same is not true for an {\it ultra-} relativistic point charge. In most of the afore-mentioned excellent textbooks, what is presented is a plausibility argument of a possible relativistic generalization from the non-relativistic formula for the energy and momentum of radiation from a point charge, giving a formula whose only significance is that it has the correct non-relativistic limit. One is left wondering that perhaps there are other relativistic formulas with an identical non-relativistic limit. We avoid such issues by presenting first in this paper a derivation of the radiant energy-momentum due to a relativistic accelerated charge, starting with the Lorentz-covariant radiant energy-momentum tensor derived earlier manifestly relativistically. A more complete, deductive {\it derivation} of this 4-momentum of radiation due to a relativistic, arbitrarily-moving point charge is also given, with inspiration from the formulation of S. Coleman \cite{coleman} of a classical charge which employs Fourier-transformed distributions. 

This beautiful formulation has the added advantage of the ability of inclusion of radiation backreaction, once again, manifestly Lorentz-covariantly. The relativistic charge equations of motion, with radiative backreaction included, immediately exhibit a pathology which forecast inherent problems of classical electrodynamics like {\it acausality}, from a completely relativistic standpoint. These issues are usually discussed in the literature in the non-relativistic limit \cite{coleman}. Our fully relativistic formulation enables us to derive, in appropriate limits the well-known Abraham-Lorentz \cite{abralor} and Landau-Lifschitz \cite{ll-ctf} equations, without the need to use any extraneous structural artifacts, like assuming the point charge to be a charged sphere of vanishingly small radius, as done in some contemporary papers (e,g, \cite{grif}).

The paper is organized as follows :  In section 2, we consider a single arbitrarily-moving point charge as the source, and obtain its potentials and field strengths thereof. Focusing on the radiative parts of the field strengths, we obtain the expression of our covariant energy-momentum tensor. We then perform the spatial integration to obtain the relativistic 4-momentum of the radiation fiuelds due to an arbitrarily moving point charge. We further go on to compute the orbital angular momentum of the radiation fields.  In the next section, this energy-momentum tensor is used in its Fourier tranformed version to derive the 4-momentum of a relativistic point charge. In section 4, we discuss the effect of including radiation reaction, and analyse the resulting equations leading to runaway solutions and/or {\it preacceleration} despite resorting to mass renormalization. This section follows the analysis of ref. \cite{coleman} fairly closely. The inadequacy of Maxwell electrodynamics is thus reviewed in this section. In this section we also rederive the Abraham-Lorentz and Landau-Lifschitz equations for a point charge directly from our results, without having to make any `model' of the point charge. We conclude briefly in section 5.  

\section{Radiation from a point charge}

\subsection{Radiant energy-momentum tensor}

The Maxwell equations for the physical potentials $A^{\m}$ are given by
\begin{eqnarray}
\Box A^{\m} &=& J^{\m} \nonumber \\
\partial_{\m} A^{\m} & =& 0~.  \label{maxeq}
\end{eqnarray}
The current density 4-vector can be easily written down for a relativistic point charge with the parametric equation of the trajectory ${\bar x}^i= {\bar x}^i(\tau) $ where $\tau$ is the proper time of the charge; it is given by
\begin{eqnarray}
J^{\m}(x) = e\int d\tau ~u^{\m}(\tau)~\delta^{(4)}(x-{\bar x}(\tau))~. \label{cur}
\end{eqnarray}
where the 4-velocity of the charge $u^{\n} \equiv d{\bar x}^{\n}/d\tau$. As is well-known \cite{ll-ctf}, \cite{jack}, substitution of this 4-current density into eqn. (\ref{maxeq}) leads to the Lienard-Wiechert 4-potentials (the subscript P has been dropped)
\begin{eqnarray}
A_{\m}= \frac{e~u_{\m}(\tau_0)}{({\bf R}(\tau_0) \cdot {\bf u(\tau_0)})}
\end{eqnarray}
where, $\tau=\tau_0$ corresponds to the proper time at which the world line of the charge just enters the past light cone of the observer. Thus, the 4-vector ${\bf R}(\tau) \equiv {\bf x} - {\bf {\bar x}}(\tau) $, giving the spacetime interval separating the charge from the observer, becomes {\it null} at $\tau_0: R^2(\tau_0) =0$. It is obvious that this defines the {\it retarded} time ${\bar t} = t - |{\vec r} - {\vec {\bar r}(\tau_0)}|$. 

The corresponding field strength tensor components can also be obtained \cite{jack}; it is easy to see that the field tensor can be decomposed into two parts, the `radiation' and `Coulomb' parts :
\begin{eqnarray}
F_{\m \n} &=& F_{\m \n}^{rad}+F_{\m \n}^{Cou} ~\label{fdec} \\
F_{\m \n }^{rad} &=& e~\frac{(R_{\m} a_{\n} - R_{\n} a_{\m}) ({\bf u} \cdot {\bf R}) + (u_{\m}R_{\n}-u_{\n}R_{\m}) ({\bf a} \cdot {\bf R})} {({\bf u} \cdot {\bf R})^{3}} \label{frad} \\ 
F_{\m \n}^{Cou} &=& e~ \frac{u^2 (u_{\m} R_{\n} - u_{\n} R_{\m})}{({\bf u} \cdot {\bf R})^{3}} ~. \label{fcou} 
\end{eqnarray}
It is to be remembered that all quantities are evaluated at $\tau=\tau_0$ which corresponds to the retarded time. One obvious distinction between these two parts above is the dependence of the `radiation' part on the 4-acceleration ${\bf a}$ of the source charge of which the `Coulomb' part is independent. Another distinction emerges if we choose a frame in which the null vector ${\bf R} = |{\vec R}|(1, {\hat R})$, where ${\hat R} \equiv {\vec R}/|{\vec R}|$ : for large $|{\vec R}|$,  $F_{\m \n}^{Cou}$ falls off as $1/|{\vec R}|^2$, hence justifying the superscript `Coulomb', while $F_{\m \n}^{rad}$ falls off only as fast as $1/|{\vec R}|$, implying that the radiation field strength survives much farther away compared to the Coulomb field strength. From this point on, therefore, we shall ignore the Coulomb field strength.

The energy-momentum tensor of the electromagnetic field is given in terms of the field strength tensor by the well-known formula \cite{ll-ctf}
\begin{equation}
T^{\m \n} =  \left[ F^{\m \r}~F^{\r}_{~\n} - \frac14 \eta^{\m \n} F^{\r \lambda}~F_{\r \lambda} \right]~. \label{enm}
\end{equation}
Substituting $F^{rad}_{ij}$ from eqn (\ref{frad}), we obtain the fully Lorentz-covariant radiation energy-momentum tensor due to an arbitrarily-moving point charge
\begin{eqnarray}
T_{rad}^{\m \n } = - e^2~\frac{R^{\m} R^{\n}}{({\bf u} \cdot {\bf R})^6}~\left[ a^2 ({\bf u} \cdot {\bf R})^2 + u^2 ({\bf a} \cdot {\bf R})^2 \right] ~. \label{rkmvu}
\end{eqnarray} 
We note that the expression (\ref{rkmvu}) has not yet appeared in standard textbooks, to the best of our knowledge. It is also clear that in the frame where ${\bf R} = |{\vec R}| (1, {\hat R})$, the radiant energy-momentum tensor components fall off at large $|{\vec R}|$ as $1/|{\vec R}|^2$, so that the angular integral of $|{\vec R}|^2 T^{0 \m}$ does not vanish on the surface of 2-sphere of very large radius, in this frame. 

The angular distribution of the radiant power due to an accelerated, relativistic point charge can be computed from (\ref{rkmvu}),
\begin{eqnarray}
\frac{d {\cal P}}{d \Omega} \equiv |{\vec R}|^2~ {\hat R}_{\alpha} ~T^{0 {\alpha}}_{rad} ~.\label{radp}
\end{eqnarray}
It is straightforward to check that this angular power distribution, expressed in terms of the 3-velocity ${\vec \beta}$ and the 3-acceleration ${\dot {\vec \beta}}$, where the `dot' corresponds to the derivative with respect to the retarded coordinate time $t$, {\it coincides} with the formula for angular power distribution due to a relativistic accelerated charge, in this frame, given in standard textbooks \cite{ll-ctf}, \cite{jack}. 

In the {\it non-relativistic} limit ${\vec \beta} \rightarrow 0$, the angular integral can be performed, yielding the Larmor formula for the power radiated by a non-relativistic, accelerated, point charge
\begin{eqnarray}
{\cal P} \simeq \frac23 ~e^2~ |{\dot {\vec \beta}}|^2 ~. \label{larm}
\end{eqnarray}

\subsection{Radiant 4-momentum due to a relativistic point charge}

It is clear that the energy-momentum of radiation due to a relativistic point charge is given by,
\begin{eqnarray}
P^{\m}_{rad} \equiv \int_{{\cal S}} d^3s ~n_{\n}~ T^{\m \n}_{rad} =  \int d^3 r ~T^{0 \m}_{rad} ~ \label{4mom}
\end{eqnarray} 
where, the first integral above is over a spacelike hypersurface ${\cal S}$ to which ${\bf n}$ is a timelike unit normal : $n^2=-1$;  here, $T^{ij}_{rad}$ is given by eqn (\ref{rkmvu}). The local conservation of the energy-momentum tensor guarantees that the energy-momentum components are independent of the choice of the spacelike hypersurface. The second equality results on choosing a frame in which ${\bf n} = (-1, {\vec 0})$. However, unlike in the non-relativistic case, the angular integral in the relativistic case, in the chosen frame is difficult to perform, to give a closed form expression for the radiant energy-momentum. In standard textbooks \cite{ll-ctf} \cite{jack}, the energy-momentum 4-vector of the radiation due to a relativistic charge is postulated by trying to `lift' the Larmor formula (\ref{larm}) to spacetime, by expressing it in terms of 4-vectors ${\bf a}$ and ${\bf u}$, such that the Larmor formula ensues in the non-relativistic limit. This, clearly is not a satisfactory derivation, since there may be other Lorentz-covariant expressions with the same non-relativistic limit. 

Using the first equality in (\ref{4mom}), we write the radiant 4-momenta as
\begin{eqnarray}
P^{\m} = -e^2 ~\int_{\cal S} d^3x ~\frac{R^{\m} ({\bf n} \cdot {\bf R}) ~\left[ a^2 ({\bf u} \cdot {\bf R})^2  + u^2 ({\bf a} \cdot {\bf R})^2 \right]} {({\bf u} \cdot {\bf R})^6} ~. \label{rkmom}
\end{eqnarray}
Recall that the spacetime vector ${\bf u}$ is timelike with the squared norm $u^2=1$ for $c=1$; if we set ${\bf n} = {\bf u}$ without loss of generality, the above expression reduces to
\begin{eqnarray}
P^{\m} = -e^2 ~\int_{\cal S} d^3x ~\frac{R^{\m} ~\left[ a^2 ({\bf u} \cdot {\bf R})^2  + ({\bf a} \cdot {\bf R})^2 \right]} {({\bf u} \cdot {\bf R})^5}
\end{eqnarray}

We may, once again, without loss of generality, choose the spacelike hypersurface ${\cal S}$ to have a tangent vector given by ${\bf a}$ at the event where the normal to ${\cal S}$ is the 4-velocity ${\bf u}$. This point corresponds to the value $\tau=\tau_0$ which defines the retarded time. Observe, then, that the null 4-vector ${\bf R}$ has the resolution at this point
\begin{eqnarray}
R^{\m} = ({\bf u} \cdot {\bf R})~u^{\m} + \frac{({\bf a} \cdot {\bf R})~a^{\m}}{a^2} + R_{\perp}^{\m} ~\label{rdec}
\end{eqnarray}
where, ${\bf u} \cdot {\bf R}_{\perp} = 0 = {\bf a} \cdot {\bf R}_{\perp}$. It follows that the pullback to the hypersurface ${\cal S}$ can be written, choosing local Euclidean coordinates on ${\cal S}$, as,
\begin{eqnarray}
R^{\alpha} |_{\cal S} = \frac{({\bf a} \cdot {\bf R})~a^{\alpha}}{a^2} |_{\cal S} + R_{\perp}^{\alpha} |_{\cal S} ~\label{rdecs}
\end{eqnarray}

Using these results, one can write:
\begin{eqnarray}
    P^{\mu}&=&-e^{2}a^{2}\int _{S}d^{3}R\frac{R^{\mu}(R_{u}^{2}-R_{a}^{2})}{R_{u}^{5}}\label{relmom2}
\end{eqnarray}
Using the properties of the delta function,we can re-write \eqref{relmom2} as:
\begin{eqnarray}
P^{\mu}&=&-e^{2}a^{2}\int d\tau \int _{S}d^{3}R\frac{(R^{2}-R_{\perp}^{2})}{R_{u}^{4}}\delta(R^{2})\{R_{u}u^{\mu}+R_{a}\frac{a^{\mu}}{(-a^{2})^{1/2}}+R_{\perp}^{\mu}\}\label{relmom3}
\end{eqnarray}
Using our choice of S, we can now write $d^{3}R=dR_{a}d^{2}R_{\perp}$ and then after noting that $\int dR_{a}R_{a}=0=\int d^{2}R_{\perp} R_{\perp}^{\mu}$, \eqref{relmom3} becomes:
\begin{eqnarray}
    dP^{\mu}&=& e^{2}a^{2}u^{\mu}d\tau \int dR_{a}d^{2}R_{\perp}\frac{R_{\perp}^{2}}{2R_{u}^{3}(R_{u}^{2}-R_{\perp}^{2})^{1/2}}\{\delta(R_{a}-\sqrt{R_{u}^{2}-R_{\perp}^{2}})+\{\delta(R_{a}+\sqrt{R_{u}^{2}-R_{\perp}^{2}})\}\label{relmom5}\\
    &=&\frac{e^{2}a^{2}u^{\mu}}{2R_{u}^{3}}d\tau \int _{R_{\perp}^{2}=0}^{R_{\perp}^{2}=R_{u}^{2}} d^{2}R_{\perp}\frac{R_{\perp}^{2}}{\sqrt{R_{u}^{2}-R_{\perp}^{2}}}\\
    &=&\frac{e^{2}a^{2}u^{\mu}}{2R_{u}^{3}}d\tau \times \frac{4}{3}R_{u}^{3}
\end{eqnarray}
Thus, we get the relativistic Larmor formula for the radiant power \cite{ll-ctf}, \cite{jack}:
\begin{eqnarray}
    P&=&\frac{dP^{\mu}}{d\tau}u_{\mu}=\frac{2}{3}e^{2}a^{2}\label{rel-lar}
\end{eqnarray}
where $a^{2}$ is now the four-acceleration squared and \eqref{rel-lar} is thus a manifestly Lorentz co-variant expression.\\
However, in this derivation, we have completely neglected the back-reaction of the radiation field on the particle without which the equation of motion of a relativistic charged particle and hence the limitations of classical electrodynamics remain unexplored. In the next section, we shall derive the complete equation of motion of the charged particle, once again in a manifestly Lorentz covariant manner.

\subsection{Radiant Orbital Angular Momentum due to a relativistic accelerated charge}

Using the fields \eqref{fdec}, \eqref{fcou} and \eqref{frad} The angular momentum radiated by a relativistic point charge can also be computed in an exact Lorentz covariant manner as follows.\\
We begin by writing the Noether-Current Density corresponding to the in variance of the action under Lorentz Transformations:
\begin{eqnarray}
    J_{\rho \mu \nu} &=& J_{\rho \mu \nu}^{orb}+J_{\rho \mu \nu}^{spin}\\
    J_{\rho \mu \nu}^{orb}&=&R_{[\rho}T_{\mu] \nu}\\
    J_{\rho \mu \nu}^{spin}&=&A_{[\rho}\partial_{\nu}A_{\mu]}-A_{[\rho}\partial_{\mu]}A_{\nu}
\end{eqnarray}
where the sub-scripts orb and spin correspond to the orbital angular momentum (conserved due to invariance under space-time transformations) and spin (conserved due to invariance under the intrinsic transformation of the fields). In this paper, we are working strictly with classical electrodynamics and hence shall restrict ourselves to OAM only. While computing $J^{orb}$, unlike the the previous sub-section, we shall include the entire energy momentum tensor and not just the radiation part \eqref{rkmvu}, for reasons that we shall subsequently clarify :
\begin{eqnarray}
    T_{\mu \nu}&=&F_{\mu \rho}F^{\rho}_{\nu}-\frac{1}{4}\eta_{\mu \nu}F_{\rho \sigma}F^{\rho \sigma}\\
    &=&T_{\mu \nu}^{col}+T_{\mu \nu}^{coup}+T_{\mu \nu}^{rad}
\end{eqnarray}
where, $T_{\mu \nu}^{col}\approx\frac{1}{r^{4}}$, $T_{\mu \nu}^{coup}\approx\frac{1}{r^{3}}$ and $T_{\mu \nu}^{rad}\approx\frac{1}{r^{2}}$\\
using these considerations, $J^{orb}$ can be written as:
\begin{eqnarray}
    J_{\rho \mu \nu}^{orb}&=&R_{[\rho}T_{\mu] \nu}^{col}+R_{[\rho}T_{\mu] \nu}^{rad}+R_{[\rho}T_{\mu] \nu}^{coup}\\
    R_{[\rho}T_{\mu] \nu}^{rad}&=& e^{2}\frac{(a^{2}(u.R)^{2}+u^{2}(a.R)^{}2)}{(u.R)^{6}}R_{[\rho}R_{\mu]}R_{\nu}=0\label{AM1}\\
    R_{[\rho}T_{\mu] \nu}^{col}&\approx&\frac{1}{r^{3}}\label{AM2}
\end{eqnarray}
From \eqref{AM1} and \eqref{AM2} it is clear that only $R_{[\rho}T_{\mu] \nu}^{coup}$ gives a non-zero contribution upon integration over all space. To evaluate $R_{[\rho}T_{\mu] \nu}^{coup}$, we first need to evaluate $T_{\m \n}^{coup}$ :
\begin{eqnarray}
    T_{\mu \nu}^{coup}&=&\eta^{\rho \sigma}\{F_{\mu \rho}^{col}F_{\nu \rho}^{rad}+F_{\nu \rho}^{col}F_{\mu \rho}^{rad}\}+\frac{1}{2}\eta_{\mu \nu}F_{\rho \sigma}^{col}F^{\rho \sigma}_{rad}\\
    &=&\frac{e^{2}}{(u.R)^{5}}\{(R_{\mu}a_{\nu}+R_{\nu}a_{\mu})(u.R)-(u_{\mu}R_{\nu}+U_{\nu}R_{\mu})(a.R)\}
\end{eqnarray}
Thus, we can now write the Orbital angular momentum density and the corresponding Noether charge:
\begin{eqnarray}
    J_{\rho \mu \nu}^{orb}&=& \frac{e^{2}R_{\nu}}{(u.R)^{5}}\{R_{[\rho}a_{\mu]}(u.R)-R_{[\rho}U_{\mu]}(a.R)\} \\
    M_{\rho \mu }^{orb}&=& e^{2}\int_{S} d^{3}S_{\nu}R^{\nu}\frac{R_{[\rho}a_{\mu]}(u.R)-R_{[\rho}u_{\mu]}(a.R)}{(u.R)^{5}}
\end{eqnarray}
Here, S is a space-like 3-surface and we can choose our co-ordinates in such a way that U is the normal and a the tangent to S since U and a are time-like vectors respectively; Thus $d^{3}S_{\nu}=d^{e}\vec{R}u_{\nu}$
Let us also re-write $R_{\mu}$ as :
\begin{eqnarray}
    R_{\mu}&=& R_{u}u_{\mu}+R_{a}\frac{a_{\mu}}{(a^{2})^{1/2}}+R_{\perp \mu}\\
     R_{u}&=& (u.R)\\
     R_{a}&=& (a.R)
\end{eqnarray}
Then $d^{3}\vec{R}$ becomes: $d^{3}\vec{R}=dR_{a}d^{2}\vec{R}_{\perp}$.
In this choice of co-ordinates, The orbital angular momentum tensor becomes:

where,
\begin{eqnarray}
    I^{(1)}_{\rho}&=& \int \frac{d^{3}\vec{R}}{R_{u}^{3}}R_{\rho}\\
    &=& \int d\tau \int\frac{dR_{a}d^{2}\vec{R}_{\perp}}{R_{u}^{2}}\{R_{u}u_{\rho}+R_{a}\frac{a_{\rho}}{(a^{2})^{1/2}}+R_{\perp \rho}\}\delta(R^{2})\\
    &=& \int d\tau u_{\rho} \int\frac{dR_{a}d^{2}\vec{R}_{\perp}}{R_{u}}\delta(R_{a}^{2}-(R_{u}^{2}-R_{\perp}^{2}))\\
    &=& \int d\tau u_{\rho} \int\frac{dR_{a}d^{2}\vec{R}_{\perp}}{R_{u}\sqrt{R_{u}^{2}-R_{\perp}^{2}}}\{\delta(R_{a}-\sqrt{R_{u}^{2}-R_{\perp}^{2}})+\delta(R_{a}+\sqrt{R_{u}^{2}-R_{\perp}^{2}})\}\\
    &=& 2\int d\tau u_{\rho} \int\frac{d^{2}\vec{R}_{\perp}}{R_{u}\sqrt{R_{u}^{2}-R_{\perp}^{2}}}\\
    &=& 2\int d\tau u_{\rho}
\end{eqnarray}
and,
\begin{eqnarray}
    I^{(2)}_{\rho}&=& \int \frac{d^{3}\vec{R}}{R_{U}^{4}}R_{a}R_{\rho}\\
    &=&\int d\tau \int\frac{R_{a}dR_{a}d^{2}\vec{R}_{\perp}}{R_{u}^{4}}\{R_{u}u_{\rho}+R_{a}\frac{a_{\rho}}{(a^{2})^{1/2}}+R_{\perp \rho}\}\delta(R^{2})\\
    &=&\int d\tau\frac{a_{\rho}}{(-a^{2})^{1/2}}\int\frac{R_{a}^{2}dR_{a}d^{2}\vec{R}_{\perp}}{R_{u}^{3}}\delta(R_{a}^{2}-(R_{u}^{2}-R_{\perp}^{2}))\\
    &=&\int d\tau\frac{a_{\rho}}{(-a^{2})^{1/2}}\int\frac{R_{a}^{2}dR_{a}d^{2}\vec{R}_{\perp}}{R_{u}^{3}\sqrt{R_{u}^{2}-R_{\perp}^{2}}}\{\delta(R_{a}-\sqrt{R_{u}^{2}-R_{\perp}^{2}})+\delta(R_{a}+\sqrt{R_{u}^{2}-R_{\perp}^{2}})\}\\
    &=& 2\int d\tau\frac{a_{\rho}}{(-a^{2})^{1/2}}\int\frac{d^{2}\vec{R}_{\perp}}{R_{u}^{3}}\sqrt{R_{u}^{2}-R_{\perp}^{2}}\\
    &=& \frac{2}{3}\int d\tau \frac{a_{\rho}}{(-a^{2})^{1/2}}
\end{eqnarray}
Subsituting these into the expression of $M_{\rho\mu}$, we get:
\begin{eqnarray}
    \frac{dM_{\rho \mu}}{d\tau}&=& 4u_{[\rho}a_{\mu]}(1+\frac{1}{3(-a^{2})^{1/2}})\label{covorb}\\
    \implies \frac{dL_{i}}{dt}&=& \frac{4}{\gamma}\epsilon_{ijk}u^{j}a^{k}(1+\frac{1}{3(-a^{2})^{1/2}})\label{ORB}
\end{eqnarray}
Eqn \eqref{covorb} is the fully covariant Angular Momentum Tensor and eqn \eqref{ORB} is the rate of angular-momentum loss by radiating relativistic point particle. These expressions are yet to appear in standard literature to the best of our knowledge and can be used to compute the rate of angular momentum loss by an arbitrarily moving relativistic charged particle.

\section{Energy-momentum of a relativistic charge revisited}

\subsection{In and Out fields}

In this section, the radiant four-momentum is computed in another manner, such that the inclusion of radiation backreaction can be carried out in a straightforward manner. Following \cite{coleman}, this approach employs {\it incoming} and {\it outgoing} fields whose use in quantum field theory is well-known. Their use in classical electrodynamics in 1960 was no doubt a novelty. The approach automatically encompasses the backreaction of radiant potentials on the source charge in a fully Lorentz-covariant manner, and leads to the physics of radiation reaction without extraneous structural assumptions on the charge. 

We shall define the In and Out fields as follows:
\begin{eqnarray*}
    A_{\mu}^{in}=A_{\mu}-A^{R}_{\mu} \\
    A_{\mu}^{out}=A_{\mu}-A_{\mu}^{adv}
\end{eqnarray*}
where $A_{\mu}^{R(A)}$ is the retarded (resp advanced) four potential. These so called "incoming" and "outgoing" fields are both solutions of the homogeneous equation $\partial_{\mu}\partial_{\mu}A_{\nu}^{in(out)}=0$.Next we state a general fact about the solutions to the inhomogeneous equation, by defining :
\begin{equation}
    A^{f}_{\mu}(t)=\int d^{3}x\{A_{\mu}(x,t)\frac{\partial f(x,t)}{\partial t}-f(x,t)\frac{\partial A_{\mu}}{\partial t} \}
\end{equation}
with $\partial_{\mu} \partial_{\mu} f=0$ . Now we state the following assertion that justify the names "incoming" and "out-going" fields: 
\begin{eqnarray}
    \lim_{t \to -\infty}A_{\mu}^{f}=A_{\mu}^{f(in)} \\
    \lim_{t \to \infty}A_{\mu}^{f}=A_{\mu}^{f(out)}
\end{eqnarray}
The proof of the assertions (38) and (39) are given in \cite{coleman}. 
Now, having justified the "In" and "Out" nomenclature, we proceed to state the boundary condition of Maxwellian electrodynamics that "All radiation comes from somewhere" i.e. 
\begin{equation}
    A_{\mu}^{in}=0
\end{equation}
It is to be noted that (40) holds only when we have accounted for all the charges in the universe.

Now, the four momentum of the the radiation field can be written as
\begin{eqnarray}
P_{\mu}=P_{\mu}(A^{out})-P_{\mu}(A^{in})=P_{\mu}(A^{out}) ~, \label{pout}
\end{eqnarray}
since $A^{in}_{\mu}=0$. Now, the radiation 4-potentials
\begin{eqnarray}
A_{\mu}=A_{\mu}^{R}=\int d^{4}x' D_{R}(x-x') J_{\mu}(x') &=& A_{\mu}^{out} +\int d^{4}x'D_{A}(x-x')J_{\mu}(x') \tilde{A}_{\mu}^{out}
\end{eqnarray}
Thus 
\begin{eqnarray}
A_{\mu}^{out}=\int d^4{x'} J_{\mu}(x') [D_{R}(x-x') -D_{A}(x-x')]=\int D(x-x') J_{\mu}(x')d^{4}x'
\end{eqnarray}
where $D_{R(A)}$ is the retarded (resp advanced) Green's function corresponding to the d'Alembertian operator, and $D$ is one-half the difference between these Green's functions. Upon Fourier tranformation,
\begin{eqnarray}
\tilde{A}_{\mu}^{out}=\tilde{D} (k) \tilde{J}_{\mu}(k)=2 \pi i \epsilon(k_{0}) \delta (k^{2})\tilde{J}_{\mu}(k) \equiv \tilde{a}_{\mu}(k) \delta (k^{2}) ~. \label{ak}
\end{eqnarray}
where $\epsilon(k^0)$ is the sign function. To compute the radiant energy-momentum, we start with the integral 
\begin{eqnarray}
t_{\mu \nu} \equiv \int d^3x ~ T_{\mu \nu}^{rad}(A^{out})
\end{eqnarray}
It is obvious that the energy-momentum $P_{\mu} \equiv t_{\mu0}$. Now, to compute $t_{\mu \nu}$, it is convenient to evaluate first 
\begin{eqnarray}
\int \partial_{\mu}A^{out}_{\lambda}\partial_{\nu}A^{out}_{\rho} d^{3}x  &=& \int  \frac{d^{4}k d^{4}k'}{(2 \pi)^{8}} k_{\mu}k'_{\nu} \tilde{a}_{\lambda}(k) \tilde{a}_{\rho}(k')\delta(k^{2}) \delta(k'^{2})  \int e^{ix_{\sigma}(k_{\sigma}+k'_{\sigma})}d^3x ~\label{dela}
\end{eqnarray}
The spatial integral is trivial, yielding  
\begin{eqnarray}
\int \partial_{\mu}A^{out}_{\lambda}\partial_{\nu}A^{out}_{\rho} d^{3}x = \int \int  \frac{d^{4}k dk'_{0}}{(2 \pi)^{5}} k_{\mu}\tilde{k}_{\nu} \tilde{a}_{\lambda}(\vec{k},k_{0}) \tilde{a}_{\rho}(\vec{-k},k'_{0})\delta((\vec{k})^{2}-k_{0}^{2}) \delta((\vec{k})^{2} -k_{0}^{'2}) e^{ix_{0}(k_{0}+k'_{0})} \label{delaa}
\end{eqnarray}
This immediately leads, upon using properties of Dirac delta functions, to perform the integral over $k^{'0}$, so that
\begin{equation}
    \int \partial_{\mu}A^{out}_{\lambda}\partial_{\nu}A^{out}_{\rho} d^{3}x = \frac{1}{2} \int \frac{d^{4}k}{(2\pi)_4}\frac{k_{\mu}k_{\nu}}{|k_{0}|}\tilde{a}_{\lambda}(k) \tilde{a}_{\rho}(-k)\delta(k^{2}) ~. \label{dela2}
\end{equation}
This implies that  
\begin{equation}
   t_{\mu \nu} = \frac{1}{2}\int \frac{d^{4}k}{(2\pi)^4}\frac{k_{\mu}k_{\nu}}{|k_{0}|}\tilde{j}_{\lambda}(k) \tilde{j}_{\lambda}(-k)\delta(k^{2}) \nonumber
\end{equation}

\subsection{Energy-momentum of relativistic charge}

Going back to position space we get
\begin{eqnarray}
t_{\mu \nu} = \frac{1}{2}\int \int j_{\lambda}(x) {j}_{\lambda}(x')d^{4}x d^{4}x'\int \frac{k_{\mu}k_{\nu}}{|k_{0}|}e^{ik_{\rho}(x_{\rho}-x'_{\rho})} \delta(k^{2})\frac{d^{4}k}{(2\pi)^4} ~. \label{enmom}
\end{eqnarray}
This enables us to compute the four momentum of the radiation field
\begin{eqnarray}
P_{\mu}= t_{\mu 0} = \frac{1}{2}\int  j_{\lambda}(x) {j}_{\lambda}(x')d^{4}x d^{4}x'\int k_{\mu} e^{ik_{\rho}(x_{\rho}-x'_{\rho})} \delta(k^{2})\epsilon (k_{0}) \frac{d^{4}k}{(2\pi)^{4}}~. \label{enm}
\end{eqnarray}
Carrying out the integration over $k$, we get
\begin{eqnarray}
P_{\mu}=\frac{1}{2}\int  j_{\lambda}(x) {j}_{\lambda}(x')\partial_{\mu}D(x-x')d^{4}x d^{4}x' ~, \label{enmo}
\end{eqnarray}
where $D(x-x')$ is the Green's function: $$D(x-x')=\int \delta(k^{2})e^{ik_{\rho}(x_{\rho}-x'_{\rho})} \frac{d^{4}k}{(2\pi)^{4}} .$$ 

Substituting the expression (\ref{cur}) for the 4-current density for a point charge, we get
\begin{eqnarray}
P_{\mu}=\frac{e^{2}}{2}\int d\tau d\tau ' u_{\lambda}(\tau)u_{\lambda}(\tau') \partial_{\bar \mu} D({\bar x}(\tau)-{\bar x}(\tau')) = \int d\tau u_{\lambda}(\tau)\int u_{\lambda}(\tau') \partial_{\bar \mu} D(w)d\tau' ~,\label{pii}
\end{eqnarray}
where $\partial_{\bar \mu} \equiv (\partial/\partial {\bar x}^\mu)$ and $w \equiv (({\bar x(\tau)} - ({\bar x}(\tau'))^2)^{1/2}$. We evaluate the integral over $\tau'$ first. Let
\begin{eqnarray}
I_{\lambda \mu} \equiv \int u_{\lambda}(\tau') \partial_{\bar \mu} D(w)d\tau' ~, \label{iil}
\end{eqnarray}
Using 
$$\partial_{\bar \mu}=\frac{{\bar x}_{\mu}}{w}\frac{d}{dw} \quad and \quad u_{\lambda}d\tau=\frac{d{\bar x}_{\lambda}}{dw}~ dw$$
We get :
$$I_{\mu \lambda}=\int^{\infty}_{0} \frac{d{\bar x}_{\lambda}}{dw}\frac{{\bar x}_{\mu}}{w}\frac{d D}{dw}~ dw. $$
Integrating by parts, we see that
\begin{equation}
      I_{\mu \lambda}= -\int^{\infty}_{0} dw D(w)~\frac{d}{dw}(\frac{{\bar x}_{\mu}}{w}\frac{d{\bar x}_{\lambda}}{dw}) 
\end{equation}

\subsection{Regularization and Mass Renormalization}

Now, the Green's function $D(w) = \delta(w^2)$ and therefore, the maximal contribution to $I_{\mu \lambda}$ must come from around $w=0$, as it should, since the radiation emanates from a single relativistic accelerated charge. To enable a Taylor expansion around $w=0$, we work with the {\it regularized} Green's function $D_{\Lambda}(w)$ defined by the requirement that $\lim_{\Lambda \rightarrow \infty} D_{\Lambda}(w) = D(w)$. A particularly suitable regularized Green's function is the Gaussian-regularized Green's function
\begin{eqnarray}
D_{\Lambda}(w) = \frac{\Lambda}{\sqrt{2\pi}}~\exp - \Lambda^2 w^2 ~. \label{gaureg}
\end{eqnarray}
Define $x'_\mu \equiv (d{\bar x}_{\mu}/dw)_{w=0}$, and likewise for the higher order derivatives, we obtain,
\begin{eqnarray}
{\bar x}_{\mu} \approx x'_{\mu} w + x''_{\mu} \frac{w^{2}}{2} + x'''_{\mu} \frac{w^{3}}{6} + \cdots ~\label{tau}
\end{eqnarray}
This implies that,
\begin{eqnarray}
I_{\mu \lambda} = \int_0^{\inf} dw~D_{\Lambda}(w)~\left[ x'_{\mu} x''_{\lambda} + \frac12 x''_{\mu}x'_{\lambda} + w \left ( \frac13 x'_{\lambda} x'''_{\mu} + x''_{\lambda} x''_{\mu} + x'''_{\lambda} x'_{\mu} \right) \right] + {\cal O}(\Lambda^{-1}) ~. \label{il} 
\end{eqnarray}
Performing the integrals and evaluating the limits, we obtain,
\begin{eqnarray}
I_{\mu \lambda}=[\Lambda(x''_{\lambda} x'_{\mu}+ \frac{x'_{\lambda} x''_{\mu}}{2} )+ \frac{1}{3} x'_{\lambda} x'''_{\mu} + x''_{\lambda} x''_{\mu} + x'_{\mu} x'''_{\lambda}] + {\cal O} (\frac{1}{\Lambda})~. \label{il2}
\end{eqnarray}

To reproduce the usual formula for the energy-momentum (\ref{rel-lar}) of a relativistic point charge, we need to convert the derivatives w.r.t. $w$ to derivatives w.r.t. the affine parameter $\tau$ evaluated at $\tau=\tau_0$. To this end, we use the relation
\begin{eqnarray}
\frac{d}{dr}|_{r=0}=(\frac{d\tau}{dr}|_{r=0})\frac{d}{d\tau}
\end{eqnarray}
to yield
\begin{eqnarray}
x'_{\mu} &=& (\frac{d\tau}{dr}|_{r=0}) \frac{dx_{\mu}}{d\tau}=(\frac{d\tau}{dr}|_{r=0}) u_{\mu} \nonumber \\
x''_{\mu} &=& (\frac{d\tau}{dr}|_{r=0})^{2} a_{\mu}+(\frac{d^{2}\tau}{dr^{2}}|_{r=0}) u_{\mu} \nonumber \\
x'''_{\mu} &=& (\frac{d\tau}{dr}|_{r=0})^{3} {\dot a_{\mu}} + 3(\frac{d^{2}\tau}{dr^{2}}|_{r=0})(\frac{d\tau}{dr}|_{r=0}) a_{\mu}+(\frac{d^{3}\tau}{dr^{3}}|_{r=0}) u_{\mu} ~, \label{taud} 
\end{eqnarray}                               
where the `dot' signifies a $\tau$ derivative. To evaluate the derivatives, we again expand r in a Taylor series in powers of $\tau-\tau_0$ about $\tau$ assuming r=0 at $\tau=\tau_0$; to simplify the notation, we redefine $\tau-\tau_0 \equiv \tau$, so that the expansion is about $\tau=0$ in powers of $\tau$. This leads to 
\begin{eqnarray}
x_{\mu}\approx 0+ u_{\mu}(0) \tau +a_{\mu}(0) \frac{\tau^{2}}{2} + {\dot a}_{\mu}(0) \frac{\tau^{3}}{6} + \cdots
\end{eqnarray}
Using this, we get
\begin{eqnarray}
r=({\bar x}_{\rho} {\bar x}_{\rho})^{\frac{1}{2}} \approx \sqrt{ \tau^{2} \left( 1-\frac{\tau^{2} a^2}{12} \right)}\approx \tau(1-\frac{\tau^{2} a^2}{24})=\tau-\frac{\tau^{3} a^2}{24}
\end{eqnarray}
From here, it is staightforward to show that
$$\frac{d\tau}{dr}|_{\tau=0}=1; \quad \frac{d^{2}\tau}{dr^{2}}|_{\tau=0}=0; \quad\frac{d^{3}\tau}{dr^{3}}|_{\tau=0}=-\frac{1}{4}(\ddot{x_{\rho}}\ddot{x_{\rho}}) $$
Substituting in (\ref{il2}), one obtains,

\begin{equation}
    I_{\mu \lambda}=[\Lambda({\dot x}_{\mu} {\ddot x}_{\lambda} +\frac{{\ddot x}_{\mu} {\dot x}_{\lambda}}{2})+ \frac{1}{4 \pi}({\dot x}_{\mu} {\dddot x}_{\lambda} +\frac{{\dddot x}_{\mu} {\dot x}_{\lambda}}{3} +{\ddot x}_{\mu} {\ddot x}_{\lambda} -\frac{{\dot x}_{\mu} {\dot x}_{\lambda}}{3} {\ddot x}_{\rho}{\ddot x}_{\rho}) ] +  {\cal O} \left(\frac{1}{\Lambda} \right)
\end{equation}

Now, using ${\bf a} \cdot {\bf u}=0$ and $\tau$-derivatives thereof
\begin{eqnarray}
  P^{\mu} = \frac{e^2}{2} \int d\tau u_{\lambda} I^{\mu \lambda} =\frac{e^2}{2}  \int d\tau \left[ \Lambda a^{\mu} + \left( \frac{{\dot a}^{\mu}}{3} - \frac23~u^{\mu} a^2 \right) \right] + {\cal O} \left(\frac{1}{\Lambda} \right) ~. \label{enmomf}  
\end{eqnarray}
The first term in eqn(\ref{enmomf}) blows up in the physical limit $\Lambda \rightarrow \infty$; this conundrum is taken care of by {\it Mass Renormalization} : one realizes that the measured mass $m$ of the particle is {\it not} the same as the mass parameter $m_0$ inserted into the Lagrangian of the particle, because of the interaction of the particle with its own electromagnetic field. The physical mass differs from the `bare' mass by the amount $e^2 \Lambda/2$. Thus, the energy-momentum of the radiation field is equal in magnitude and opposite in sign to the energy-momentum {\it lost} by the particle. This implies that 
\begin{eqnarray}
P^\mu &=& \int d\tau~m~a^\mu \nonumber\\
m &=& m_0 + \frac{e^2}{2}~\Lambda ~\label{renm}
\end{eqnarray}
It is clear that this procedure eliminates the $ \Lambda$-dependent part of the radiant energy-momentum (\ref{enmomf}), so that, realizing that, of the remaining terms, the first term is a total divergence and therefore discarded, one is left with the expression,
\begin{eqnarray}
dP^\mu = - \frac23 e^2 ~ u^\mu ~ a^2~ d\tau = - \frac23 e^2 ~a^2 d{\bar x}^\mu \label{relmof}
\end{eqnarray}
which is identical to the expression (\ref{rel-lar}).

\section{Equation of motion of charge}

\subsection{Abraham-Lorentz Equation}

Along its world line, the radiant charge loses energy, resulting in a deceleration, given by 
\begin{eqnarray}
\frac{dP_{\m}}{d\tau} =  m {a_{\m}} = e F_{\m\n} u^{\n} + \frac{e^2}{3}~\left[   {\dot a_{\m}} - 2 u_{\m} a^2 \right]~, \label{eom}
\end{eqnarray}
where, the first term on the {\it rhs} is the relativistic Lorentz force on the charge due to its own radiation field, the frist term in the square-bracket is due to radiation backreaction, while the second is loss of energy due to radiation. Thus, the formalism in this section includes the effect of radiation backreaction on the source charge in the relativistic case, in contrast to the formulation in the previous section where this was ignored. 

Recall that the charged particle follows a timelike trajectory with $u^2=-1$ and ${\bf u} \cdot {\bf a} =0$, signifying that the acceleration spacetime vector is {\it spacelike}. Notice however that there is a problem with these basic tenets, when we contract both sides of (\ref{eom}) with $u^i$, which immediately leads to $a^2 = 0$, upon using the relation $a^2 + {\bf u} \cdot {\dot {\bf a}}=0$ ! Thus, eqn (\ref{eom}) is inconsistent with our earlier contention that the spacetime acceleration is a spacelike 4-vector, or alternatively, that the charge moves along a timelike trajectory. In other words, because of the effects of radiation on the source charge, including loss of energy and also radiation backreaction, the particle is driven to move towards {\it superluminal propagation} ! This, if true, is a violation of causality, which might appear as a preacceleration, but is manifest relativistically here.

Indeed, the {\it Abraham-Lorentz} equation of motion \cite{abralor} results immediately, if one takes the non-relativistic limit of eqn (\ref{eom}) :
\begin{eqnarray}
\frac{dP_{\m}}{d\tau}={\cal F}_{(rad) \m}\approx \frac{2e^{2}}{3} {\dot a}_{\m} ~. \label{ablor}
\end{eqnarray}
Notice that in arriving at eqn(\ref{ablor}), no assumption regarding attributing any structure (like a sphere of vanishing radius) has been necessary, in contrast to certain recent assays \cite{grif}. The result is of course a non-relativistic approximation to the relativistic result (\ref{eom}), and inherits the consistency issues of that equation, in that it leads to a violation of causality in the form of a {\it preacceleration} \cite{coleman}. 

\subsection{Landau-Lifschitz Proposal}

The key feature of the Landau-Lifschitz formulation \cite{ll-ctf} is the the modification of eqn (\ref{eom}) by extra terms so that it now reads
 \begin{eqnarray}
\frac{dP_{\mu}}{d\tau} =  m {a_{\mu}} = e F_{\mu \nu} u^\nu + g_\mu~, \label{eomll}
\end{eqnarray}
where, the radiation reaction 4-force $g_i$ is now {\it constrained} to obey the relation ${\bf g} \cdot {\bf u} =0$. For this auxiliary terms have to be added to  the second term of eqn (\ref{eom}); 
\begin{eqnarray}
g_\mu = \frac{e^2}{3}~   {\dot a_{\mu}} + \alpha~u_\mu ~\label{llrar}
\end{eqnarray}
the 4-scalar $\alpha$ is now determined by the requirement that the full ${\bf g}$ must now obey the constraint ${\bf g} \cdot {\bf u} =0$, yielding the final result \cite{ll-ctf}
\begin{eqnarray}
\frac{dP_{\m}}{d\tau} =  m {a_{\m}} = e F_{\m \n} u^{\n} + \frac{2e^2}{3}~\left[   {\dot a_{\m}} +  u_{\m} a^2 \right]~. \label{eomllf}
\end{eqnarray}
It is easy to verify that the pathology discussed in the last subsection, observed by first contracting (\ref{eom}) by the 4-velocity, disappears when the same procedure is repeated for eqn (\ref{eomllf}). Thus, at this relativistically covariant level, there is no prospect of any violation of causality. However, it is also clear that there is no possible manner in which the extra terms, by which equations (\ref{eom}) and (\ref{eomllf}) differ, can be generated by utlizing the ambiguities of mass renormalization, whereby the `counter-term' added to the bare mass can be altered by terms that remain finite in the limit that the ultraviolet cutoff goes to infinity. What one needs are terms proportional to the four-velocity rather than the four acceleration, and these cannot be generated by tweaking the counterterm. 

The issue of the Landau-Lifschitz proposal, thus, is one of interpretation. In a sense, the Landau-Lifschitz work evades any discussion of the {\it physical} origin of this extra term, even though it does identify this extra term as the one that is necessary to avoid the conundra mentioned in the last subsection. As discussed in \cite{coleman}, \cite{soph}, it is unlikely that such a term can be autonomously derived from the basic tenets of classical Maxwell electrodynamics.

\section{Discussion}

The main purpose of this revisit of classical electrodynamics of an arbitrarily-moving point charge has been to exhibit its in-built special relativistic features, and to show how these lead to its essential results in a straightforward manner, including its inherent pathologies. 

Starting with the Lienard-Wiechert potentials for a relativistic, accelerated point charge, we have determined the corresponding field strengths and shown that they separate into a `radiative' and a `Coulomb' part. The radiative field strenghts combine into a rather simple-looking radiant energy-momentum tensor expressed in terms of the 4-velocity and the 4-acceleration of the particle, and the spacetime distance of the charge from the observer. `Retarded time' also emerges from this relativistic formulation, as it must. To the best of our knowledge, such a covariant energy-momentum tensor has not appeared in the literature. From this the radiant energy-momentum and orbital angular momentum have also been derived directly, as also more rigorously from Fourier space analysis. The inclusion of radiation backreaction has been shown to lead to certain inherent lacunae of classical Maxwell electrodynamics. Even though it is clear what modifications to the theory must be made to eliminate these lacunae, what remains unknown is the absence of a clear mechanism as to how these terms may arise within a classical framework. 

We have not been able to provide a satisfactory resolution of the lacunae of  Maxwell electrodynamics within a classical framework. They appear to be rather generic, and not something that minor modification of the theory can hope to achieve. What we have done is to show that a fully Lorentz-invariant, gauge invariant  approach may be the clearest approach to Maxwell electrodynamics where formal methodological encumbrances can be completely dispensed with, thereby focusing on the key conceptual conundra that are inherent in the theory. However, in most applications, these conceptual issues can be sidestepped, because radiation reaction effects are negligible in situations of practical importance. In such situations, Maxwell's great creation dazzles in brilliance within our approach where special relativity is intrinsic to its formulation. This is one aspect where we think we have made a value addition to extant textbook literature.

\section{Acknowledgments}

We acknowledge useful discussions with R. Koley.


\begin{thebibliography}{99}
\bibitem {mr} P. Majumdar and A. Ray, {\it Maxwell Electrodynamics in terms of Physical Potentials}, to appear (2019).
\bibitem {einst} A. Einstein, {\it Zur Elektrodynamik bewegter K\"orper}, Ann. der. Phys. {\bf 17} (1905) 37, reprinted as {\it On the Electrodynamics of Moving Bodies} in {\it The Principle of Relativity}, Dover (1952).  
\bibitem {ll-ctf} L. D. Landau and E. M. Lifschitz, {\it Classical Theory of Fields}, Pergamon Press (1975).
\bibitem {jack} J. D. Jackson, {\it Classical Electrodynamics}, Wiley Eastern (1974).
\bibitem {coleman} S. Coleman, {\it Classical Electron Theory from a Modern Standpoint}, Monograph published by Rand Corporation (1961), reprinted in {\it Electromagnetism}, ed. D. Teplitz, Springer, Boston (1982) . 
\bibitem {abralor} M. Abraham, {\it Theorie der Elektrizit ...}, Vol. II (Teubner, Leipzig ,1905); H. A. Lorentz, {\it The Theory of Electrons ...}, (Leipzig, New York,1909).
\bibitem {rohr} F. Rohrlich, Phys. Lett. A 283 (2001) 276; Phys. Rev. E 77 , 046609 (2008).
\bibitem {grif} D. J. Griffiths, T.C. Proctor, D.F. Schroeter, {\it Abraham-Lorentz Versus Landau-Lifshitz} Am. J. Phys., Vol. 78, No. 4, April 2010.
\bibitem {bm-epjc} S. Bhattacharjee and P. Majumdar, {\it Gauge-free Coleman-Weinberg Potential}, Eur. Phys. Jou. {\bf C73} (2013) 2348. e-Print arXiv:1302.7272
\bibitem{soph}  H. Spohn {\it The critical manifold of the Lorentz-Dirac equation} 2000 EPL 50 287
\end{thebibliography}
\end{document}